\documentclass[journal,draftcls,onecolumn,12pt,twoside]{IEEEtran}
\IEEEoverridecommandlockouts
\usepackage{cite}
\usepackage{amsmath,amssymb,amsfonts}
\usepackage[linesnumbered,ruled,vlined]{algorithm2e}
\usepackage{graphicx}
\usepackage{epstopdf}
\usepackage{textcomp}
\usepackage{xcolor}
\usepackage{multirow}
\usepackage{soul}
\usepackage{array}
\usepackage{booktabs}
\def\BibTeX{{\rm B\kern-.05em{\sc i\kern-.025em b}\kern-.08em
    T\kern-.1667em\lower.7ex\hbox{E}\kern-.125emX}}
\begin{document}
%\titlespacing{\chapter}{0pt}{-2pt}{2pt}
%\title{\large \hspace{-1.5em}Asynchronous Federated Learning Based Mobility-aware \hspace{-1.5em} \par Caching in Vehicular Edge Computing }
\title{\huge Asynchronous Federated Learning Based Mobility-aware Caching in Vehicular Edge Computing }
%\titleformat*{\chapter}[display]{\large}
%\title{Asynchronous Federated Learning Based Mobility-aware Caching in Vehicular Edge Computing }
%\author{Wenhua Wang, Yu Zhao and Qiong Wu,~\IEEEmembership{Member,~IEEE}

%\thanks{This work was supported in part by the National Natural Science Foundation of China under Grant No. 61701197, in part by the open research fund of State Key Laboratory of Integrated Services Networks under Grant No. ISN23-11,  in part by the 111 Project under Grant No. B12018. \emph{(Corresponding author: Qiong Wu)}

%Wenhua Wang, Yu Zhao and Qiong Wu are with School of Internet of Things Engineering, Jiangnan University, Wuxi 214122, China, and also with the State Key Laboratory of Integrated Services Networks (Xidian University),  Xi'an 710071, China (Email: wenhuawang@stu.jiangnan.edu.cn, yuzhao@stu.jiangnan.edu.cn, qiongwu@jiangnan.edu.cn).
%}
%}
\author{\IEEEauthorblockN{Wenhua Wang$^{1,2}$, Yu Zhao$^{1,2}$, Qiong Wu$^{1,2}$, Qiang Fan$^{3}$, Cui Zhang$^{4}$ and Zhengquan Li$^{1,5}$}
	\IEEEauthorblockA{\small(1. School of Internet of Things Engineering, Jiangnan University, Wuxi 214122, China)}
	\IEEEauthorblockA{\small(2. State Key Laboratory of Integrated Services Network(Xidian University),Xi'an 710071, China)}
	\IEEEauthorblockA{\small(Email: \{wenhuawang, yuzhao\}@stu.jiangnan.edu.cn, qiongwu@jiangnan.edu.cn)}
	\IEEEauthorblockA{\small(3. Qualcomm, San Jose CA 95110 USA, China)}
	\IEEEauthorblockA{\small(Email: qiangfan29@gmail.com)}
	\IEEEauthorblockA{\small(4. Banma Network Technology Co., Ltd., Shanghai 200000, China)}
	\IEEEauthorblockA{\small(Email: zc351340@alibaba-inc.com)}
	\IEEEauthorblockA{\small(5. Jiangsu Future Networks Innovation Institute, Nanjing 211111, China)}
	\IEEEauthorblockA{\small(Email: lzq722@jiangnan.edu.cn)}
	
	\thanks{This work was supported in part by the National Natural Science Foundation of China (No. 61701197), in part by the open research fund of State Key Laboratory of Integrated Services Networks (No. ISN23-11),  in part by the 111 Project (No. B12018),  in part by the Future Network Scientific Research Fund Project (FNSRFP-2021-YB-11).
	}	
	}	
\maketitle
%(FNSRFP-2021-YB-11)
\begin{abstract}
Vehicular edge computing (VEC) is a promising technology to support real-time applications through caching the contents in the roadside units (RSUs), thus  vehicles can fetch the contents requested by vehicular users (VUs) from the RSU within short time. The capacity of the RSU is limited and the contents requested by VUs change frequently due to the high-mobility characteristics of vehicles, thus it is essential to predict the most popular contents and cache them in the RSU in advance. The RSU can train model based on the VUs' data to effectively predict the popular contents. However, VUs are often reluctant to share their data with others due to the personal privacy. Federated learning (FL) allows each vehicle to train the local model based on VUs' data, and upload the local model to the RSU instead of data to update the global model, and thus VUs' privacy information can be protected. The traditional synchronous FL must wait all vehicles to complete training and upload their local models for global model updating, which would cause a long time to train global model. The asynchronous FL updates the global model in time once a vehicle's local model is received. However, the vehicles with different staying time have different impacts to achieve the accurate global model. In this paper, we consider the vehicle mobility and propose an Asynchronous FL based Mobility-aware Edge Caching (AFMC) scheme to obtain an accurate global model, and then propose an algorithm to predict the popular contents based on the global model. Experimental results show that AFMC outperforms other baseline caching schemes.

\end{abstract}


\vspace{-0.25cm}
\section{Introduction}
\vspace{-0.15cm}
%\IEEEPARstart{W}{ith} the advancement of the internet of vehicles (IoV), caching technology facilitates the development of the real-time vehicular applications \cite{Liuchen2021}. However, traditional caching technology is not sufficient for the stringent transmission delay requirements of the vehicular users (VUs) to fetch the contents, as the vehicles need to fetch the requested contents from the remote cloud. Vehicular edge computing (VEC) can significantly alleviate the above problem \cite{Javed2021}, which allows vehicles to fetch requested contents directly from adjacent road side unit (RSU) that caches the contents.
\IEEEPARstart{W}{ith} the advancement of the internet of vehicles (IoV), caching technology facilitates the development of the real-time vehicular applications \cite{Liuchen2021,Wu2022}. Vehicles typically fetch the contents requested by vehicular users (VUs) from a macro base station (MBS) connected with a cloud to support the vehicular applications. However, the cloud is far from the vehicles, thus the stringent delay requirement to fetch contents may not be satisfied. Vehicular edge computing (VEC) is a promising technology to significantly reduce the delay to fetch contents, which consists of a MBS connected with a cloud and a road side unit (RSU) deployed at the edge \cite{Zhu2021}. The MBS can cache all available contents due to its large storage capacity while the RSU can retrieve contents from the MBS and cache them. The capacity of the RSU is limited, thus it only caches part of the available contents. VUs in the VEC can fetch contents directly from the RSU, thus satisfying the delay requirement\cite{Javed2021}.

For the traditional caching schemes, the contents are cached based on the previously requested contents. However, vehicles in the VEC enter and leave the coverage of a RSU frequently owing to the high-mobility characteristics, which incurs the frequent changes of contents requested by VUs. Thus the traditional caching scheme cannot ensure that the RSU accurately caches VUs' requested contents, which would result in that vehicles cannot fetch contents from the RSU successfully. It is essential to predict the most popular contents and cache them in the RSU in advance. With the assistance of machine learning (ML), the RSU can train a model through extracting features from VUs' data to effectively predict the popular contents \cite{Yan2019}. However, owing to privacy issue, VUs are often reluctant to share their data with each other, which results in difficulties for RSU to collect data to train the model.

Federated learning (FL) allows each vehicle to train the local model based on its VUs' data, and then upload the local model to the RSU for the global model updating, and thus FL can significantly protect VUs' privacy information. Some works have studied the FL based caching in the VEC. In \cite{Yu2017}, Yu \emph{et al.} proposed a mobility-aware proactive edge caching scheme based on FL which allows multiple vehicles to participate in training the global model to predict popular contents in VEC, thus the growing demand for computationally intensive and latency-sensitive vehicular applications can be met. In \cite{Chilukuri2020}, Chilukuri \emph{et al.} proposed an adaptive cache allocation scheme for edge caching based on FL in a dynamic and resource constrained vehicular network. In \cite{Cui2015}, Cui \emph{et al.} designed a FL-based compression algorithm aided by blockchain to predict the popular contents in VEC. In \cite{Lu2020}, Lu \emph{et al.} proposed a FL based scheme consisting of intelligent data transformation and collaborative data leakage detection to achieve dynamic content caching in VEC. However, the above methods adopted the synchronous FL to design caching schemes, where all vehicles have to train and upload their local models before the RSU aggregates all local models to update the global model, which would cause a very large time to train global model. In \cite{Xie2019}, Xie \emph{et al.} proposed an asynchronous FL to reduce the training time through updating the global model once a uploaded local model is received. However, the RSU may receive a local model uploaded from a vehicle which has small staying time in the coverage area of the RSU, thus the contents required by the VUs of the vehicle may become outdated quickly, which may further deteriorate the accuracy of the global model. Hence, it is critical to consider the vehicle mobility in designing the asynchronous FL in VEC to improve the accuracy of the global model. To the best of our knowledge, there is no work considering the vehicle mobility in asynchronous FL in VEC, which motivates us to conduct this work.

In this paper, we propose an Asynchronous FL based Mobility-aware Edge Caching (AFMC) scheme to predict accurate popular contents in the VEC. We first design an asynchronous FL framework considering the mobility of vehicles to improve the accuracy of the global model. Then we adopt the autoencoder (AE) to predict the popular contents based on the global model.

%Motivated by the above discussion, it encourages us to devise an asynchronous FL
%considering the mobility of vehicles to predict accurate popular contents in VEC. We design an Edge Caching scheme based on Asynchronous FL (ECAF) in the highly dynamic VEC, in order to predict the most popular contents and cache them in the RSUs in advance. First, we design asynchronous FL framework considering the mobility characteristics of vehicles to improve the accuracy of the global model. To predict the accurate popular contents, the autoencoder (AE) based on the global model is adopted to extract the hidden features of the local data to improve the cache efficiency.

%Motivated by the above discussion, it encourages us to design an Mobility-Aware Edge Caching scheme based on Asynchronous FL (MECAF) in the highly dynamic VEC. \st{In this scheme, if a vehicle completes updating the local model and uploads it to the RSU, the global model is immediately updated at once without waiting for other vehicles to upload their models, thus the accuracy of the global model can be improved. ???}. First, we design an asynchronous FL framework considering the mobility characteristics of vehicles to improve the accuracy of the global model. Second, to predict the accurate popular contents, the autoencoder (AE) based on the global model is adopted to extract the hidden features of the local data to improve the cache efficiency.

The rest of this paper is organized as follows. Section \ref{sec2} briefly describes the system model. Section \ref{sec3} presents the implementation of the proposed AFMC scheme in detail. We present some simulation results in \ref{sec4}, and then conclude them in Section \ref{sec5}.
\vspace{-0.1cm}
\begin{figure}
\vspace{-0.25cm}
\center
\includegraphics[scale=0.5]{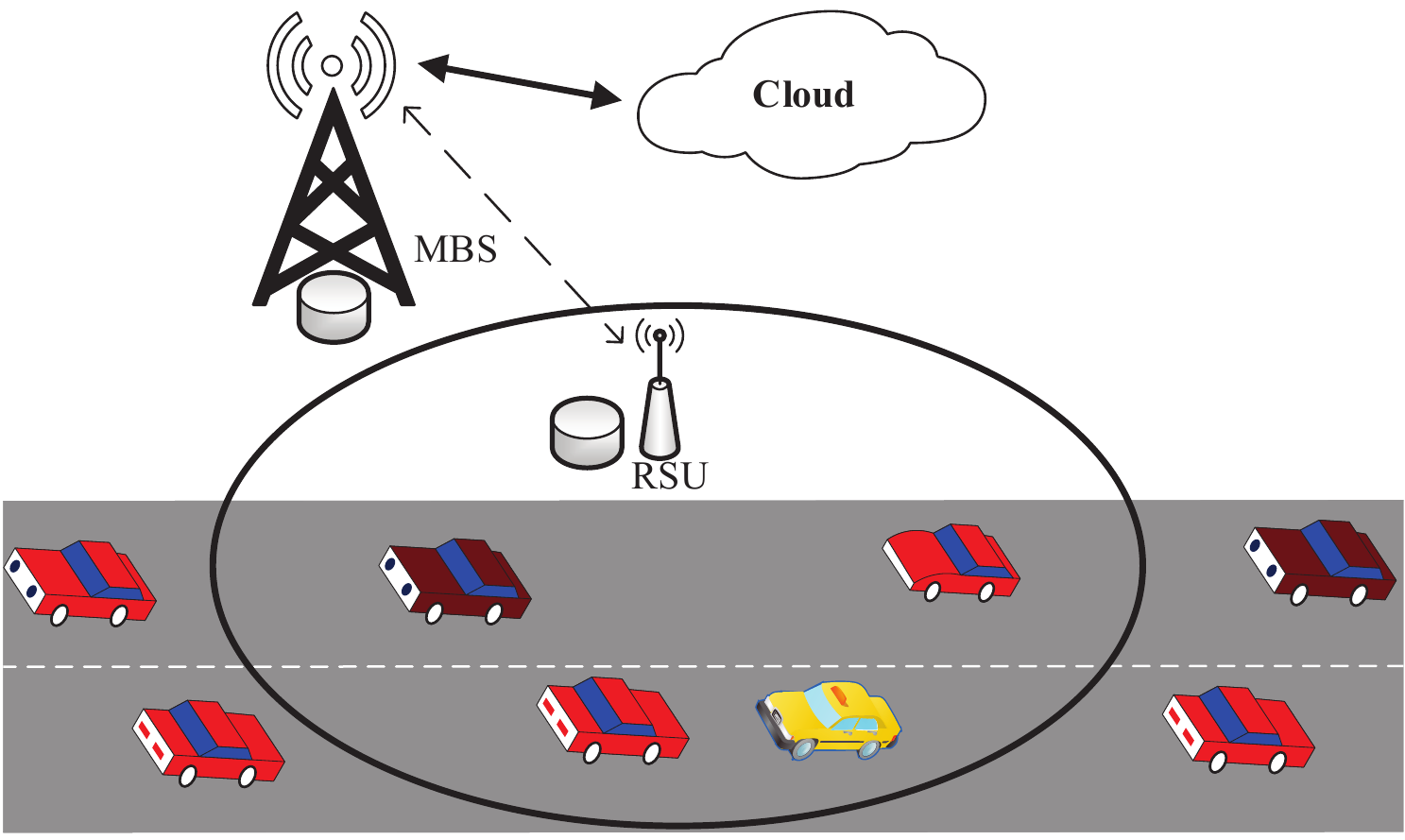}
\caption{VEC scenario}
\label{fig1}
\vspace{-0.8cm}
\end{figure}

\vspace{-0.25cm}
\section{System Model}
\label{sec2}
\vspace{-0.15cm}
 We consider a  three-tier VEC framework shown in Fig. \ref{fig1}, which comprises a macro base station (MBS) connected with a cloud, a RSU in the coverage area of the MBS and some vehicles driving in the coverage area of the RSU. The top tier is the MBS which caches all available contents. The middle tier is the RSU deployed at the edge which only caches part of contents. The bottom tier is the vehicles. Each vehicle carries serval VUs and caches the local data, where each data is a vector reflecting the VUs' personal information and ratings for all available contents. The rating for a content may be 0, which represents that the VU is uninterested in this content or the VU has not requested this content earlier. The local data are partitioned into the training set and testing set according to a certain percentage.

The VUs in each vehicle generate the information about the requested contents while each vehicle collects the requested information and sends it to the RSU to fetch the VUs' requested contents. If the RSU has the requested contents, the vehicle can fetch the contents from the RSU successfully; otherwise it has to fetch contents from the MBS.

\vspace{-0.35cm}
\section{Asynchronous Federated Learning based Mobility-aware Caching scheme}
\label{sec3}
%\vspace{-0.3cm}
In this section, we introduce the proposed AFMC. We first design a mobility-aware asynchronous FL algorithm to train an accurate global model. Then a content prediction algorithm is proposed to predict popular contents based on the trained global model.
\vspace{-0.25cm}
\subsection{Mobility-aware asynchronous FL}
\vspace{-0.2cm}
%There will execute $R^{max}$ rounds of training on the designed asynchronous FL based on AE which consists of an encoder and a decoder. Each round comprises the following steps:
The asynchronous FL algorithm executes $R^{max}$ rounds of training. Each round $r$ comprises the following steps:

\subsubsection{Vehicle Selection}

%Considering the mobility characteristics of vehicles on the road, only vehicles with sufficient staying time in the coverage area of the RSU can guarantee the accomplishment of a complete asynchronous FL process, so
Denote $V_{i}^{r}$ as the $i$th-vehicle in round $r$ within the coverage area of the RSU. In each round $r$, the vehicles with sufficient staying time to participate in asynchronous FL are first selected. Thus for each selected vehicle $V_{i}^{r}$, its staying time within the coverage of the RSU $T_{r,i}^{staying}$ should be larger than the sum of the average training time $T_{training}$ and the inference time $T_{inference}$, i.e., $T_{r,i}^{staying}>T_{training}+T_{inference}$, where $T_{r,i}^{staying}$ is calculated as
\vspace{-0.25cm}
\begin{equation}
\vspace{-0.25cm}
T_{r,i}^{staying}=\left(L_{s}-P_{i}^{r}\right) / U_{i}^{r}.
\label{eq1}
\end{equation}here $L_{s}$ is the coverage range of the RSU, $P_{i}^{r}$ is the distance from $V_{i}^{r}$ to the entrance of the RSU and $U_i^r$ is the velocity of $V_{i}^{r}$ which is generated by a truncated Gaussian distribution within the velocity limit $\left[U_{\min }, U_{\max }\right]$,
\vspace{-0.25cm}
\begin{equation}
\begin{small}
\vspace{-0.25cm}
f({U_i^r}) = \left\{ \begin{aligned}
\frac{{{e^{ - \frac{1}{{2{\sigma ^2}}}{{({U_i^r} - \mu )}^2}}}}}{{\sqrt {2\pi {\sigma ^2}} (erf(\frac{{{U_{\max }} - \mu }}{{\sigma \sqrt 2 }}) - erf(\frac{{{U_{\min }} - \mu }}{{\sigma \sqrt 2 }}))}},\\
{U_{min }} \le {U_i^r} \le {U_{max }},\\
0 \qquad \qquad \qquad \qquad \quad otherwise.
\end{aligned} \right.
\label{eq2}
\end{small}
\end{equation}where $\sigma^{2}$ is the variance, $\mu$ is the mean and $erf(x)$ is the Gauss error function.

\subsubsection{Model Download and Local Training}

The selected vehicles first download the global model $\omega^{r}$ that is aggregated at the end of the previous round from the RSU. Particularly, the RSU generates the global model based on AE for the first round. Then each selected vehicle performs $e$ iterations to update their local models. For each iteration $k$, the selected vehicle $V_{i}^{r}$ randomly retrieves some training data $n_{i,k}^{r}$ from the training set and feeds each training data $x$ $(x \in n_{i,k}^{r})$ into the AE to reconstruct $x$, where the reconstructed data, denoted as $\hat{x}$, reflects the hidden features of the data $x$. The loss function of $V_{i}^{r}$'s local model is then calculated as
\vspace{-0.4cm}
\begin{equation}
\vspace{-0.25cm}
f(\omega_{i,k}^{r})=\frac{1}{\left|n_{i, k}^{r}\right|}\sum_{k=1}^{\left|n_{i, k}^{r}\right|} l_{i}\left(\omega_{i,k}^{r};x\right),
\label{eq3}
\end{equation}where $l_{i}\left(\omega_{i, k}^{r} ; x\right)=(x-\hat{x})^{2}$ is the reconstruction error, $\left|n_{i, k}^{r}\right|$ is the number of training data for $V_{i}^{r}$ in iteration $k$ and $\omega_{i, k}^{r}$ is the local model of $V_{i}^{r}$ in iteration $k$.

To improve the convergence of the asynchronous FL, the deviation between $\omega_{i, k}^{r}$ and $\omega^{r}$ is incorporated into the loss function as a regularization term, i.e.,

%Then the regularization ?? loss function is calculated to improve the convergence of the asynchronous FL through incorporating the deviation between $\omega_{i, k}^{r}$ and $\omega^{r}$ as a regularization term in Eq. \eqref{eq3}, i.e.,
\vspace{-0.25cm}
\begin{equation}
g_{i}\left(\omega_{i,k}^{r}\right)=f_{i}\left(\omega_{i,k}^{r}\right)+\frac{\rho}{2}\left\|\omega^{r}-\omega_{i,k}^{r}\right\|^{2},
\label{eq4}
\vspace{-0.2cm}
\end{equation}where $\rho$ is the regularization parameter.
%However, \hl{vehicles may fail to upload the updated local model to RSU due to delayed staying time in the previous round}, thus conversely deteriorate the performance of the global model. To avoid this problem, an aggregated local gradient is employed considering the delayed local gradient $\nabla g_{i}^{d}$ by using the decay coefficient $\beta$ to balance the delayed local gradient and the current local gradient, defined as follows:

However, vehicles may fail to upload the their local models to the RSU, referred to as the delayed local model, due to the long training time in the previous round. These delayed local models will be uploaded to the RSU in the later rounds to update global model, which conversely deteriorates the performance of the global model, thus the local gradient should be aggregated considering the gradient of the delayed local models $\nabla g_{i}^{d}$, i.e.,
\vspace{-0.2cm}
\begin{equation}
\nabla \zeta_{i, k}^{r}=\nabla g\left(\omega_{i, k}^{r}\right)+\beta \nabla g_{i}^{d},
\label{eq5}
\vspace{-0.25cm}
\end{equation} where $\nabla g_{i}\left(\omega^{r}_{i,k}\right)$ is the gradient of $g_{i}(\omega^{r}_{i,k})$ and $\beta$ is the weighting coefficient. Then the local model is updated as
\vspace{-0.25cm}
\begin{equation}
\omega_{i, k+1}^{r}=\omega^{r}_{i,k}-\eta_{l}^{r} \nabla \zeta_{i, k}^{r},
\label{eq6}
\vspace{-0.25cm}
\end{equation} where $\eta_{l}^{r}$ is the local learning rate in round $r$ (note that
$\eta_{l}^{r}=\eta_{l} \cdot \max \{1, \log (r)\}$) $\eta_{l}$ is the initial value of local learning rate. Then $V_{i}^{r}$ executes iteration $k+1$ to update the local model. $V^{r}_{i}$ keeps updating the local model until the number of iterations reaches $e$. Then the local model is updated as $\omega_{i}^{r}$.

\subsubsection{Upload Updated Model and Asynchronous Aggregation}

Once $V^{r}_{i}$ finishes the local model updating, it will upload $\omega_{i}^{r}$ to the RSU to update the global model. Considering the vehicles with different staying time have different effect on the accuracy of the global model, the RSU would update the global model $\omega^{r}$ as

\vspace{-0.25cm}
\begin{equation}
\omega^{r}=\omega^{r-1}+\frac{d_{i}^{r}}{d^{r}} \chi_{i} \omega_{i}^{r},
\vspace{-0.2cm}
\label{eq7}
\end{equation}where $d^{r}_{i}$ and $d^r$ represent the local data size in $V^{r}_{i}$ and the total local data size of the selected vehicles, respectively, $\chi_{i}=T_{r,i}^{passing}/{T_{r,i}^{total}}$ is the weight for $V_{i}^{r}$, where $T_{r,i}^{passing} = P^{r}_{i} / U^{r}_{i}$ indicates the time after the vehicle enters the coverage area of the RSU, and $T_{r,i}^{total} = L^{r}_{i} / U^{r}_{i}$ indicates the total time of the vehicle staying in the RSU coverage area. Thus we have $\chi_{i}= {P^{r}_{i}/L^{r}_{i}}$. Note that $\chi_{i}$ is a large value if the vehicle stays in the coverage of the RSU for a longer time and thus has a higher impact on the global model.

Then the RSU will send the global model to all vehicles for the next round of updates. The RSU keeps updating the global model the number of rounds reaches $R^{max}$; Then a more efficient global model $\omega^{r}$ is achieved. After that, each vehicle within the coverage area of the RSU downloads and adopts the trained global model to predict popular contents. The specific steps about popular content prediction are described in detail in subsection B.
\vspace{-0.25cm}
\subsection{Content Popularity Prediction}
\vspace{-0.25cm}
In this subsection, we describe the popular content prediction algorithm in the following steps.

\subsubsection{Data Preprocessing}
Each vehicle $V^{r}_{i}$ abstracts the local data from the testing set to form a rating matrix $\boldsymbol{R}_{i}^{r} \in \mathbb{R}^{n\times m} $, where the rows of the matrix represent $n$ VUs and the columns of the matrix represent the ratings for $m$ contents. Nevertheless, the value 0 in the matrix represents that the VU is uninterested in this content or the VU has not requested this content, thus the value 0 will incur difficulties to in predicting popular contents. To solve this problem, each vehicle adopts the trained global model based on AE to reconstruct the rating matrix $\boldsymbol{R}_{i}^{r}$, the reconstructed the rating matrix $\hat{\boldsymbol{R}}_{i}^{r} \in \mathbb{R}^{n\times l} (l < m)$ contains few zero elements and thus it can reflect the hidden features of data.

\subsubsection{Cosine Similarity}
Each vehicle $V^{r}_{i}$ abstracts the personal information matrix $\boldsymbol{Q}_{i}^{r} \in \mathbb{R}^{n\times w} $ from the testing set and then merge it with $\hat{\boldsymbol{R}}_{i}^{r}$ to form the matrix $\boldsymbol{H}_{i}^{r} \in \mathbb{R}^{n\times(l+w)}$, where the rows of the matrix represent $n$ VUs and the columns of the matrix represent $w$ VUs' information. Define the first $s$ VUs with largest number of non-zero elements in $\boldsymbol{R}_{i}^{r}$ as the active VUs. Then the similarity between any two active VUs $a$ and $b$ is measured by the cosine similarity, i.e.,
\vspace{-0.25cm}
\begin{equation}
\begin{array}{r}
\operatorname{sim}_{a, b}^{r, i}=\cos \left(\boldsymbol{H}_{i}^{r}(a,:), \boldsymbol{H}_{i}^{r}(b,:)\right) \\
=\frac{\boldsymbol{H}_{i}^{r}(a,:) \cdot \boldsymbol{H}_{i}^{r}(b,:)^{T}}{\left\|\boldsymbol{H}_{i}^{r}(a,:)\right\|_{2} \times\left\|\boldsymbol{H}_{i}^{r}(b,:)\right\|_{2}},
\end{array}
\label{eq8}
\vspace{-0.25cm}
\end{equation} where $\boldsymbol{H}_{i}^{r}(x,:)$ denotes the vector of VU $x$ in matrix $\boldsymbol{H}_{i}^{r}$, and $\left\|x\right\|_{2}$ is the 2-norm of $x$.
\subsubsection{Interested Contents}
In each vehicle $V^{r}_{i}$, VUs with the $K$ largest similarities of each active VU are selected as the active VU's neighboring VUs. Then the ratings of $m$ contents evaluated by the $K$ neighboring VUs of $s$ active VUs are expressed as  $\hat{\boldsymbol{H}}_{K} \in \mathbb{R}^{(s \cdot K)\times m}$, where the rows of the matrix represent $K$ neighboring VUs of $s$ active VUs and the columns of the matrix represent the ratings for $m$ contents. Then each vehicle counts the number of the nonzero value of each content in $\hat{\boldsymbol{H}}_{K}$ as the content popularity of the content and selects the contents with $F_{c}$ largest content popularity as its predicted interested contents.

%Then a content with a nonzero value in $\hat{\boldsymbol{H}}_{K}$ is refered as the VU's interested content. Then the VU counts the number of interested contents

%selects the contents with $F_{c}$ largest content popularity as its predicted interested contents, \hl{where the counted number of a content is referred to as the content popularity of the content}.

%and the vectors of all these neighboring VUs for every active VU in $\boldsymbol{R}_{i}^{r}$ are constituted into a matrix $\hat{\boldsymbol{H}}_{K} \in \mathbb{R}^{(s \cdot K)\times m}$, which represents the ratings of $m$ contents evaluated by the $K$  neighboring VUs of $s$ active VUs. $V^{r}_{i}$ selects the contents with $F_{c}$ largest content popularity as the predicted interested contents.
\subsubsection{Popular Contents}
Each vehicle sends its predicted interested contents to the RSU, the RSU compares uploaded contents to select the contents with the $F_c$ largest content popularity as the popular contents. 
%The process of the AFMC scheme is shown in Algorithm \ref{al1}, where $\mathbb{V}^{r}$ is the set of vehicles within the coverage area of the RSU, $C_{r}$ is popular contents in $r$-$th$ round and $C_{i}$ is predicted interested contents of $V^{r}_{i}$ in $r$-$th$ round.

\vspace{-0.25cm}
\section{Simulation Results}
\vspace{-0.2cm}
\label{sec4}
In this section, we conduct simulation verify the effectiveness of the proposed AFMC scheme. The performance values of different schemes are obtained through averaging the results conducted in five simulation experiments.

% We first describe the experimental configuration, and then provide the numerical results compared to other five baseline schemes to validate the performance of the AFMC scheme.

\vspace{-0.25cm}
\subsection{Simulation Setup}
\vspace{-0.2cm}
The simulation tool is Python 3.8. The coverage range of RSU is 1km. The dataset we used is MovieLens 1M \cite{Harper2016}, which contains 1 million ratings for 3,883 movies from 6,040 users, as well as users' personal information including gender, age, occupation and postcode. Each user obtain local data randomly from the MovieLens 1M dataset. We randomly allocate $80\%$ data of the local data as the training set, while the remaining data as the testing set. A part of movies are randomly sampled from testing set as VU's requested contents.
\vspace{-0.25cm}
\subsection{Performance Evaluation}
\vspace{-0.2cm}
Cache efficiency is employed to reflect the probability that vehicles fetch requested contents from the RSU successfully to evaluate the performance of AFMC.
%In the edge caching scheme in VNs we have discussed, cache hit ratio is considered as a good metric to evaluate the network performance. In this paper, the cache hit rate is calculated for each RSU as follows:
\vspace{-0.25cm}
\begin{equation}
\begin{small}
\text { Cache efficiency }=\frac{\text { cache hits }}{\text { cache hits }+\text { cache misses }} \times 100 \%,
\label{eq9}
\end{small}
\end{equation}
where a cache hit indicates that a requested content is cached in the RSU and thus the vehicle can fetch requested content from the RSU successfully, while a cache miss means that a requested content is not cached in the RSU.

We compare our proposed AFMC with five baseline caching schemes described below:
\begin{itemize}
	\item Random: $N$ contents are randomly selected from all available contents to cache in the RSU.
	\item Thompson Sampling: Beta function with parameters $\alpha$ and $\beta$ is taken as a probability density function to generate the probabilities that contents are selected to be cached in RSU within $[0,1]$, where cache hits and cache misses are taken as $\alpha$ and $\beta$, respectively. Then $N$ contents with the highest probabilities are selected to be cached in the RSU.
	\item N-$\epsilon$-greedy: $N$ contents with the largest numbers of requests are selected with probability $1-\epsilon$ and $N$ contents are randomly selected from the all available contents with probability $\epsilon$. In our simulation, $\epsilon= 0.1$.
	\item FedAVG: The typical synchronous FL scheme where the RSU needs to wait for all vehicles to upload their local models and then adopts the weighted average method to update the global model.
	\item AFC: Asynchronous FL based caching scheme without considering the high-mobility characteristics of vehicles, i.e., $\chi_{i}$ in Eq. \eqref{eq7} has not been considered.
	
	%m-$\epsilon$-greedy scheme is a simple extension of $\epsilon$-greedy scheme, which selectes the highest reward contents with the probability $\epsilon$ and randomly selectes contents with the probability $1-\epsilon$ for caching. In our simulation, $\epsilon= 0.3$.
\end{itemize}

%\begin{figure}[b]
%	\center
%	\includegraphics[scale=0.32]{figure/methods_cs_ce.eps}
%	\caption{Cache efficiency under different cache capacities}
%	\label{fig2}
%\end{figure}
% \usepackage{multirow}

\begin{table*}[ht]
	\caption{Cache efficiency under different cache capacities.}
	\label{tab1}
	\footnotesize
	\centering
	\begin{tabular}{|c|cccccccc|}
		\hline
		\multirow{2}{*}{Scheme} & \multicolumn{8}{c|}{Cache capacity}                                           \\
		%\specialrule{0em}{1pt}{1pt}
		\cline{2-9}
		& 50      & 100     & 150     & 200     & 250     & 300     & 350     & 400     \\  \hline
		%\specialrule{0em}{1pt}{1pt}
		\textbf{AFMC}                    & 11.01\% & 18.17\% & 24.17\% & 29.48\% & 34.15\% & 38.35\% & 42.16\% & 45.66\% \\
		%\specialrule{0em}{1pt}{1pt}
		\textbf{AFC}                     & 10.82\% & 18.06\% & 24.02\% & 29.20\% & 33.94\% & 38.19\% & 42.06\% & 45.56\% \\
		%\specialrule{0em}{1pt}{1pt}
		\textbf{FedAVG}                  & 10.87\% & 18.05\% & 24.00\% & 29.03\% & 34.02\% & 38.31\% & 42.14\% & 45.59\% \\
		%\specialrule{0em}{1pt}{1pt}
		\textbf{Random}                  & 1.28\%  & 2.64\%  & 3.85\%  & 5.18\%  & 6.84\%  & 7.30\%  & 8.64\%  & 9.95\%  \\
		%\specialrule{0em}{1pt}{1pt}
		\textbf{Thompspon Sampling}      & 3.90\%  & 9.29\%  & 14.46\% & 19.14\% & 23.68\% & 27.34\% & 30.59\% & 33.99\% \\
		%\specialrule{0em}{1pt}{1pt}
		$N-\epsilon$-\textbf{greedy}              & 10.04\% & 16.82\% & 22.43\% & 27.44\% & 31.92\% & 35.88\% & 39.57\% & 42.92\% \\ \hline
	\end{tabular}
\end{table*}

Table. \ref{tab1} shows the cache efficiency of different caching schemes under diverse cache capacities. The vehicle density is set as $10$ vehicles/km. It is seen that the cache efficiency of all schemes increases with the cache capacity increasing. This is because that more contents are cached when the cache capacity is large and the vehicles can fetch the requested contents with high probability. It also can be seen that AFMC scheme is superior to all other schemes. In addition, the random and thompson sampling scheme which those do not predict popular contents are worsen than AFMC and N-$\epsilon$-greedy scheme. It is because that AFMC scheme uses AE to extract hidden features of data and thus can predict popular contents efficiently. Meanwhile,  N-$\epsilon$-greedy scheme only caches the requested contents with the largest numbers of requests without extracting the hidden features of the data, and thus its cache efficiency is lower than that of AFMC scheme. In addition, Table. \ref{tab1} also shows that the caching efficiency of AFMC scheme is higher than those of AFC and FedAVG scheme. It is because that AFMC scheme considers the mobility characteristics of vehicles to update the global model once a vehicle's local model is received.
%The results show that our proposed AFEC scheme is better than other baseline caching schemes, ie., Random, m-$\epsilon$-greedy and Thompson Sampling. The cache hit radio of Random scheme is worst, our proposed AFEC scheme is better than other caching schemes because that Random and Thompson Sampling schemes can not learn the historical requested contents, and m-$\epsilon$-greedy although can learn the historical requested contents but not consider the contextual information of VUs. With the cache spaces from $50$ to $500$ contents, the cache hit ratio of different caching schemes increases.

\begin{figure}
\vspace{-0.76cm}
	\center
	\includegraphics[scale=0.45]{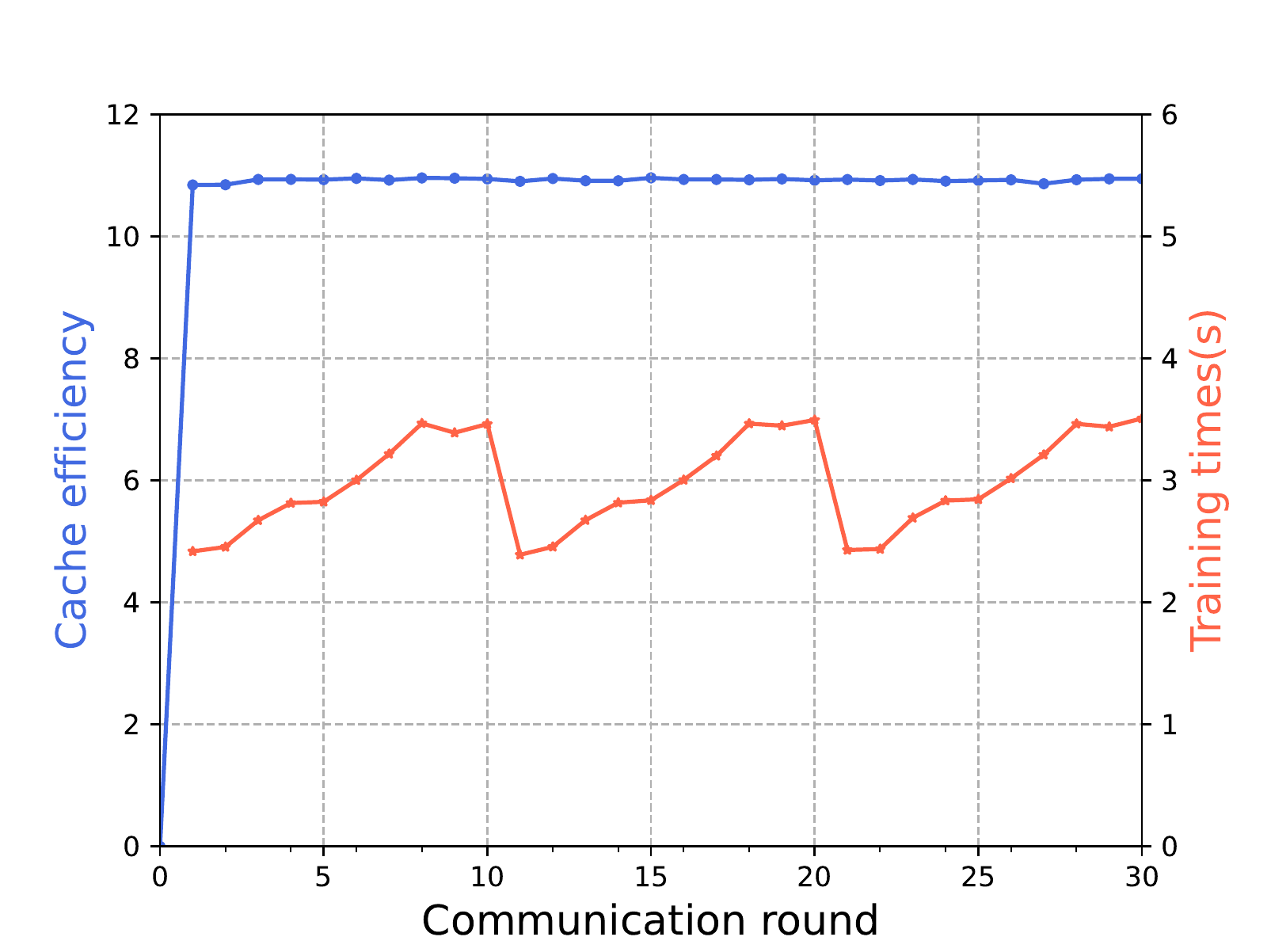}
	\vspace{-0.4cm}
	\caption{Cache efficiency and training time of AFMC}
	\label{fig3}
\vspace{-0.8cm}
\end{figure}

Fig. \ref{fig3} depicts the cache efficiency and rounds of the AFMC scheme against training time. The vehicle density is $10$ vehicles/km and the cache capacity is $50$. We can see that the cache efficiency always maintains stability around $11\%$, which demonstrates the stability of the cache performance of AFMC scheme in dynamic VEC scenarios. We can also see that the training time of the AFMC scheme for each round has a periodicity of 10 rounds. It is because that $10$ vehicles have various distributions of data and the local model can be trained fast if the data size is small. The RSU first aggregates the local model of the vehicle with the least data size and the training time gradually increases until the vehicle with the largest data completes the aggregation. Then the vehicle with the smallest data size data begins to upload the model again.
%which resulting in the training time with a periodicity of 10 communication rounds

%Fig. \ref{fig3} depicts the cache hit ratio and communication rounds against training time of our proposed AFEC scheme, in this case the number of VUs is $10$. The cache hit radio is $10.78\%$ when the communication round is $1$, and maintains stability around $11\%$ until the communication round is $30$, which demonstrates the stability of the cache performance of our proposed AFEC scheme for global model trained with multiple communication rounds in dynamic VN scenarios. Fig. \ref{fig3} also shows that the training time of asynchronous FL for each communication round is about $2.5$s, which is quite short and similarly shows that our proposed AFEC scheme is well adapted to dynamic VN scenarios.

\begin{figure}
\vspace{-0.76cm}
	\center
	\includegraphics[scale=0.45]{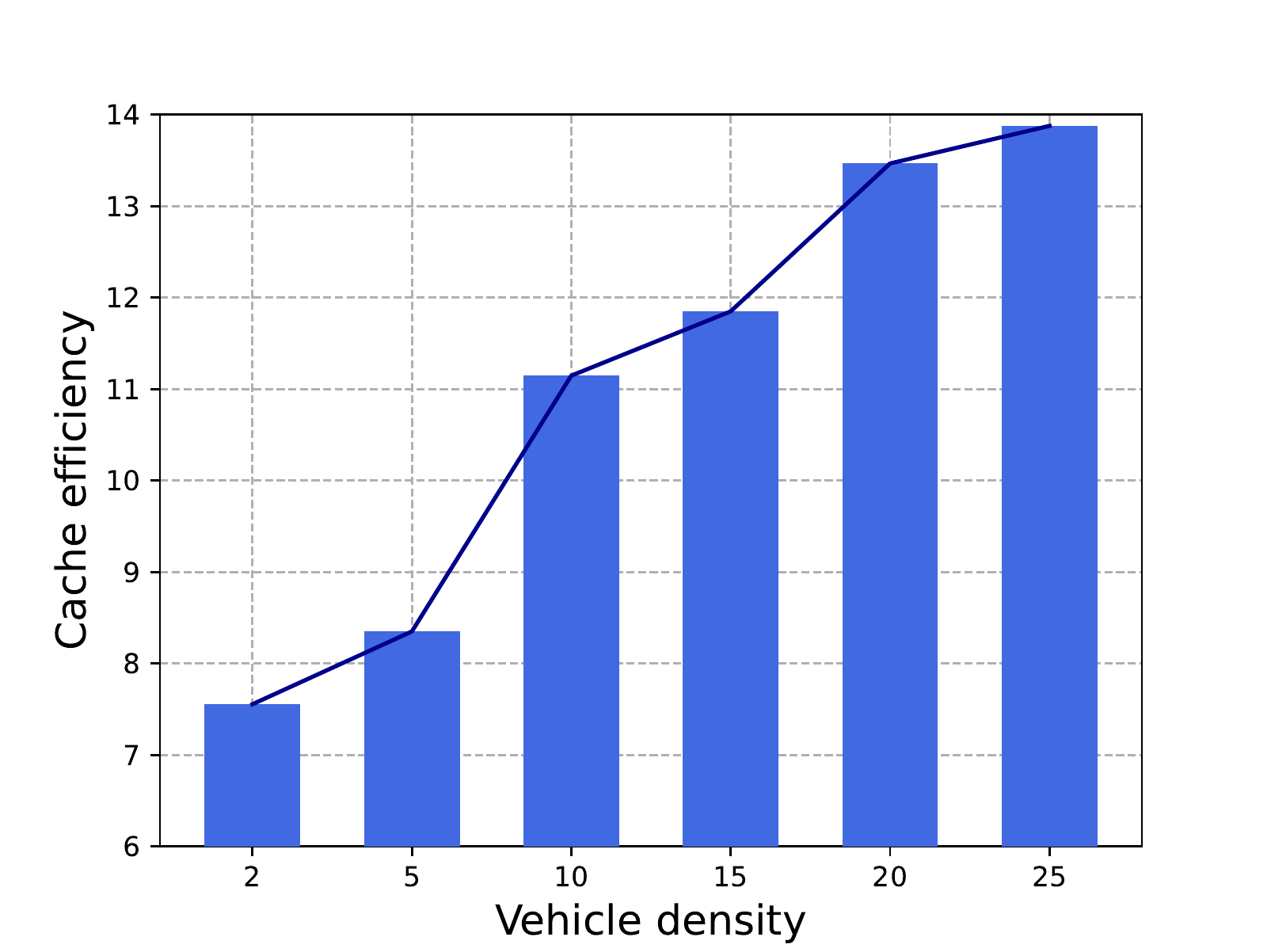}
	\vspace{-0.4cm}
	\caption{Cache efficiency of AFMC under different vehicle densities}
	\label{fig4}
\vspace{-0.5cm}
\end{figure}

Fig. \ref{fig4} depicts the cache efficiency of the AFMC scheme under different vehicle densities when the cache capacity of RSU is $50$. The cache efficiency of AFMC scheme increases from $7.55\%$ to $13.58\%$ when the number of vehicle density increases from $2$ to $25$ vehicles/km. In other words, the caching performance of the AFMC scheme will increase as more vehicles enter the coverage area of the RSU. It is because that the global model can be trained more accurately with more data when the number of vehicles increases.

\begin{figure}
	\center
	\includegraphics[scale=0.45]{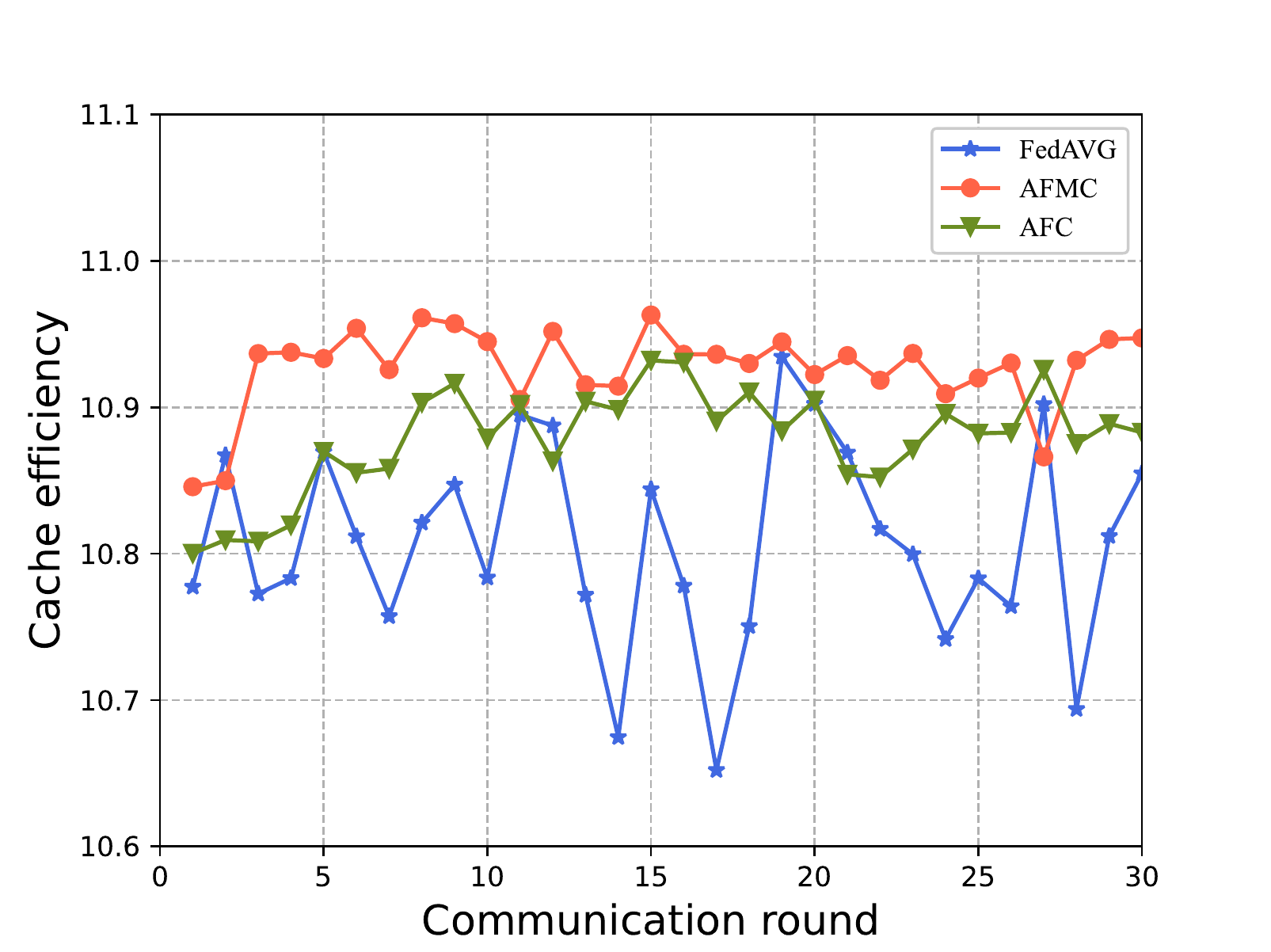}
	\vspace{-0.4cm}
	\caption{Cache efficiency under different rounds}
	\label{fig5}
\vspace{-0.7cm}
\end{figure}

Fig. \ref{fig5} compares the cache efficiency of the AFMC scheme with that of the FedAVG and AFC scheme under different rounds, where the number of vehicles is $10$ and the cache capacity of RSU is $50$. It is seen that the cache efficiency of the AFMC scheme is higher than AFC scheme. This is because the AFMC scheme considers the vehicles' mobility characteristics, and improves the accuracy of the global model. Moreover, the caching efficiency of FedAVG scheme is worsen than those of other schemes. It is because that FedAVG scheme must wait for all vehicles to upload their local models before aggregating the global model; The accuracy of the global model will be reduced if at least a vehicle has not uploaded the local model before they leave the RSU coverage. Besides, the caching efficiency of FedAVG scheme fluctuates drastically since it doesn't consider the mobility characteristics of vehicles.

\vspace{-0.2cm}
\section{Conclusions}
\vspace{-0.1cm}
\label{sec5}
In this paper, we have considered the vehicle mobility and proposed  an  AFMC caching scheme to improve the cache efficiency. We first proposed an asynchronous FL algorithm to obtain an accurate global model, and then proposed an algorithm to predict the popular contents based on the global model. Numerical results show that AFMC outperforms other baseline caching schemes. The conclusions can be summarized as follows:
\begin{itemize}
	\item AFMC scheme considers vehicles' mobility characteristics to select vehicles to participate in asynchronous FL training, which can improve the accuracy of global model.
	\item AFMC scheme greatly reduces the training time by aggregating a single vehicle's local model in each round.
\end{itemize}
\vspace{-0.2cm}

\end{document}